\documentclass{jpsj2}
\usepackage{graphicx}

\def\runauthor{K. Kudo and T. Deguchi}

\title{Level Statistics of XXZ Spin Chains with Discrete Symmetries:
Analysis through Finite-size Effects}
\author{
Kazue \textsc{Kudo} and
Tetsuo \textsc{Deguchi}$^1$}
\inst{
Graduate School of Humanities and Sciences, Ochanomizu University, 
2-1-1 Ohtsuka, Bunkyo-ku, Tokyo 112-8610\thanks{Present address:
Department of Applied Physics, Osaka City University,
3-3-138 Sugimoto, Sumiyoshi-ku, Osaka 558-8585; E-mail address:
kudo@a-phys.eng.osaka-cu.ac.jp} \\
$^1$Department of Physics, Ochanomizu University, 
2-1-1 Ohtsuka, Bunkyo-ku, Tokyo 112-8610\thanks{E-mail address:
deguchi@phys.ocha.ac.jp}}
\recdate{\today}
\abst{
Level statistics is discussed for  XXZ spin chains 
with discrete symmetries for some values of the 
next-nearest-neighbor (NNN) coupling parameter. 
We show how the level statistics of the finite-size systems 
depends on the NNN coupling and the XXZ anisotropy, which should 
reflect competition among quantum chaos, 
integrability and finite-size effects.
Here discrete symmetries play a central role in our analysis.  
Evaluating the level-spacing distribution, 
the spectral rigidity and the number variance, 
we confirm the correspondence between  
non-integrability and Wigner behavior in the spectrum. 
We also show that non-Wigner behavior appears due to mixed symmetries
and finite-size effects in some nonintegrable cases.
}
\kword{level statistics, XXZ spin chains, discrete symmetries,
finite-size effects}

\begin{document}
\sloppy
\maketitle

\section{\label{sec:intro}Introduction}

Statistical properties of energy levels have been studied 
for various physical systems in terms of the random matrix theory (RMT).  
For quantum systems, the RMT analysis has been applied to characterize
quantum chaos and to investigating the integrability of a system. For
quantum spin systems, we adopt a definition of integrability by the
Bethe ansatz: an integrable model is exactly solvable by the Bethe
ansatz. Since a pioneering work~\cite{Montam}, the following conjecture
has been widely accepted: 
If a given Hamiltonian is integrable by the Bethe ansatz, 
the level-spacing distribution should be  
described by the Poisson distribution: 
\begin{equation}
 P_{\rm Poi}(s) = \exp (-s).
 \label{eq:Poisson}
\end{equation} 
If it is  nonintegrable, the level-spacing distribution should be 
given by the Wigner distribution, i.e. the Wigner surmise for the
Gaussian orthogonal ensemble (GOE):
\begin{equation}
 P_{\rm Wig}(s) = \frac{\pi s}{2} \exp \left( - \frac{\pi s^2}{4}
                                           \right).
 \label{eq:Wigner}
\end{equation}
In principle, the conjecture is valid only for thermodynamically large
systems. 
Furthermore, there is no theoretical support for the conjecture for
quantum systems. 
However, we shall show that it is practically effective for
finite-size quantum systems.
In fact, the above conjecture has been 
numerically confirmed for many quantum spin systems 
such as correlated spin 
systems~\cite{Montam,Hsu,Poil,Pals,AM,NNN,rabson} and disordered spin
systems~\cite{Berkovits,Georgeot,Avishai,Santos,random}. 
In the Anderson model of disordered systems, $P_{\rm Poi}(s)$ and
$P_{\rm Wig}(s)$ characterize the localized 
and the metallic phases, respectively.~\cite{Shklov}

It is important to study  
statistical properties of energy levels  
for XXZ spin chains,    
which are related to various important quantum spin chains 
as well as classical lattice models in two dimensions. 
Let us consider a spin-$\frac12$ XXZ spin chain on $L$ sites
with next-nearest-neighbor (NNN) interaction  
\begin{equation}
\mathcal{H} = J_1 \sum^L_{j=1}(S^x_j S^x_{j+1} + S^y_j S^y_{j+1}
 + \Delta_1 S^z_j S^z_{j+1}) 
 + J_2 \sum^L_{j=1}(S^x_j S^x_{j+2} + S^y_j S^y_{j+2}
 + \Delta_2 S^z_j S^z_{j+2}),
\label{eq:NNN}
\end{equation}
where $S^{\alpha}=(1/2)\sigma^{\alpha}$ and 
$(\sigma^x,\sigma^y,\sigma^z)$ are the Pauli matrices; the periodic boundary
conditions are imposed. 
 The Hamiltonian~(\ref{eq:NNN}) is nonintegrable when the NNN coupling 
 $J_2$ is nonzero, while it is 
integrable when $J_2$ vanishes. Here we also note that 
it coincides with the NNN coupled Heisenberg chain when $\Delta_1=\Delta_2=1$.
When  $J_2$ vanishes the system becomes 
the integrable XXZ spin chain, 
which is one of the most important 
integrable quantum spin chains. Here,   
quantum integrability should lead to Poisson behavior as  
the characteristic behavior of level statistics. 
When $J_2$ is nonzero, 
the characteristic behavior of level statistics should be given by 
Wigner behavior. Here we note that the NNN interaction gives rise to  
 frustration among nearest neighboring and next-nearest 
 neighboring spins, which should  
lead to some chaotic behavior in the spectrum.~\cite{Nakamura}  
In a previous research, however, unexpected behavior of level-spacing 
distributions has been found for NNN coupled
XXZ spin chains.~\cite{NNN} Robust non-Wigner behavior has been seen, 
although the NNN coupled chains are nonintegrable. 
The non-Wigner behavior of level-spacing distributions 
appears particularly when total $S^z$ ($S^z_{\rm tot}$)$=0$, 
and is roughly given by 
the average of $P_{\rm Poi}(s)$ and $P_{\rm Wig}(s)$.
Similar non-Wigner behavior has been observed 
for a circular billiard when the angular momentum $L_z=0$, and 
for an interacting two-electron system with the Coulomb interaction in a 
quantum billiard when $L_z=0$.~\cite{CB} 
Here we should note that Wigner behavior has been discussed 
in ref.~\citen{rabson} 
for some XXZ spin chains in sectors of $S^z_{\rm tot} \ne 0$.

In this paper we show that finite-size effects and discrete symmetries
are important for analyzing the level statistics of XXZ spin chains.
We show explicitly how the characteristic 
property of level statistics of the XXZ spin chains 
depends on the NNN coupling, the XXZ anisotropy and the system size.
It is nontrivial to confirm the conjecture for finite XXZ spin chains
which have the integrable line and some higher symmetrical points in the
parameter space.
Numerically we discuss that the characteristic behavior of level statistics 
is determined through competition among quantum chaos, 
integrability and finite-size effects.
We confirm the correspondence between  
non-integrability and Wigner behavior in the spectrum.  
We also discuss why the unexpected non-Wigner behavior has appeared for
$S^z_{\rm tot}=0$ in ref.~\citen{NNN}. 
We explicitly consider two aspects such as mixed symmetry 
and some finite-size effects. 
Furthermore, in various cases we show under what conditions 
non-Wigner behavior appears due to the two aspects. 
It seems to be rare that such various cases of non-Wigner behavior have
been completely understood.

There is another motivation for the present research: 
another unexpected behavior of level-spacing distribution has been found 
for the integrable XXZ chain in ref.~\citen{NNN}.  The level-spacing 
distribution $P(s)$ has shown a novel peak at $s=0$ for the anisotropy
parameter $\Delta_1=0.5$. The appearance of the peak is consistent 
with the $sl_2$ loop algebra symmetry which appears only 
for special values of $\Delta_1$. Here we do not consider $\Delta_2$ 
since $J_2=0$.  
Let us introduce 
a parameter $q$ through the relation $\Delta_1=(q+1/q)/2$. 
When $q$ is a root of unity, the integrable XXZ Hamiltonian 
commutes with the $sl_2$ loop algebra.~\cite{DFM}  
Here the loop algebra is an infinite-dimensional Lie algebra, 
and the dimensions of some degenerate 
eigenspaces increase exponentially with respect to  
the system size.~\cite{FM,Deguchi} Thus the degenerate multiplicity of 
the non-Abelian symmetry can be extremely large. Furthermore, 
the $sl_2$ loop algebra is closely related to  
Onsager's algebra~\cite{Uglov}
through which the Ising model was originally solved for 
the first time.   
The XXZ spin chains are thus closely related to the most important families 
of integrable systems through the integrable point, while they 
are also extended into nonintegrable systems quite naturally. 
We may therefore expect that the RMT analysis  
of the XXZ spin chains is important  
in discussing level statistics 
for other quantum systems that have some connection to 
an integrable system.

The organization of this paper is the following. In
$\S$~\ref{sec:method}, we recall some aspects of numerical procedure 
for level statistics. In particular, we explain 
desymmetrization of the XXZ Hamiltonian. 
We also remark that in the paper we consider only 
the XY-like region where $|\Delta| \le 1$ for $\Delta=\Delta_1$ and 
$\Delta=\Delta_2$.  
In $\S$~\ref{sec:NNN}, we show how 
competition among 
quantum chaos, quantum integrability and finite-size effects  
appears in the level statistics of the NNN coupled 
XXZ spin chains.     
We evaluate level-spacing distributions, 
spectral rigidities, and number variances,  
and confirm the RMT correspondence between  
non-integrability and Wigner behavior in the spectrum. 
We solve the non-Wigner behavior reported in ref.~\citen{NNN}
and show how the behavior of level statistics changes due to mixed symmetry
and finite-size effects. 
We also show that the characteristic behavior of level statistics does
not depend on the energy range.
This makes remarkable contrast to the level statistics
of spinless fermions evaluated in the low energy spectrum at $1/3$
filling,~\cite{Zotos} as well as to that of the Anderson model evaluated
at edge regions of spectrum.  
 In $\S$~\ref{sec:integrable} we discuss level statistics 
for a special case of the integrable XXZ spin chain where 
it has the $sl_2$ loop algebra symmetry. We observe that 
there remains many spectral degeneracies in the sector of $S^z_{\rm tot}=0$  
even after desymmetrization with respect to spin reversal symmetry. 
Finally, we give conclusions in $\S$~\ref{sec:conclusions}.

\section{\label{sec:method} Numerical Procedure}

Let us discuss desymmetrization of the Hamiltonians of the XXZ spin chains. 
When performing calculation on level statistics, 
one has to separate the Hamiltonian matrices into some sectors;
in each sector, the eigenstates have the same quantum numbers.  
The NNN coupled 
XXZ chains are invariant under spin rotation around the $z$-axis, 
translation, reflection, and spin reversal. Therefore we
consider quantum numbers for the total $S^z$ ($S^z_{\rm tot}$), the
total momentum $K_{\rm tot}$, the parity, and the spin
reversal. 
However the total momentum $K_{\rm tot}$ is invariant under reflection
only when $K_{\rm tot}=0$ or $\pi$. Thus the desymmetrization according
to parity is performed only when $K_{\rm tot}=0$ or $\pi$.
Similarly, $S^z_{\rm tot}$ is invariant under spin reversal only when 
$S^z_{\rm tot}=0$. Thus the desymmetrization according to spin reversal
is performed only when $S^z_{\rm tot}=0$.

It is convenient to use a momentum-based form for the 
Hamiltonian when we calculate  eigenvalues of the NNN coupled
chains. To obtain the form, we perform 
the Jordan-Wigner and the Fourier transformations on the original spin 
 Hamiltonian. Some details are explained in Appendix~\ref{sec:JW}. 
To calculate the eigenvalues, we use standard numerical methods, which
are contained in the LAPACK library. 

To find universal statistical properties 
of the Hamiltonians, one has to 
deal with unfolded eigenvalues instead of raw eigenvalues. The unfolding
method is detailed in refs.~\citen{NNN} and \citen{random}.

To analyze spectral properties, in this paper, we calculate three
quantities: level-spacing distribution $P(s)$, spectral
rigidity $\Delta_3(l)$, and number variance $\Sigma^2(l)$. The
level-spacing distribution is the probability function $P(s)$ of 
nearest-neighbor level-spacing $s=x_{i+1}-x_{i}$, where $x_i$'s are 
unfolded eigenvalues. The level-spacing distribution is calculated over
the whole spectrum of unfolded eigenvalues unless we specify the range. 
 The spectral rigidity is given by 
\begin{equation}
 \Delta_3(l)=\left\langle \frac1{l}\min_{a,b}
\int^{\varepsilon_0 +l/2}_{\varepsilon_0-l/2} [N_u(\varepsilon)-a
\varepsilon -b]^2 d\varepsilon \right\rangle_{\varepsilon_0},
\label{eq:Delta3}   
\end{equation}  
where $N_u(\varepsilon)=\Sigma_i \theta(\varepsilon - \varepsilon_i)$ is
the integrated density of unfolded eigenvalues and $\langle
\rangle_{\varepsilon_0}$ denotes an average over 
$\varepsilon_0$. The average is done on the whole spectrum except about
15 levels on each side of the spectrum. The expression of $\Delta_3(l)$ 
gives the least square deviation of 
$N_u(\varepsilon)$ from the best fit straight line in an interval
$l$. The number variance is given by 
\begin{equation}
 \Sigma^2(l)=\left\langle \left[ N_u\left(\varepsilon_0+\frac{l}2\right) -
N_u\left(\varepsilon_0-\frac{l}2 \right) - l  \right]^2 
\right\rangle_{\varepsilon_0} ,
\label{eq:Sigma2}
\end{equation}
where $\langle
\rangle_{\varepsilon_0}$ denotes an average over
$\varepsilon_0$.~\cite{Meyer2} The average is done on the whole spectrum
except about
10 levels on each side of the spectrum.

We calculate the spectrum 
for the 18-site chains with NNN couplings. 
The matrix size is given by the following: 
$1387 \times 1387$ for $S^z_{\rm tot}=0$ and $K_{\rm 
tot}=0$ (Here desymmetrization is performed except for spin reversal); 
$1364 \times 1364$ for $S^z_{\rm tot}=0$ and 
$K_{\rm tot}=2\pi /L$, where $L$ is the number of sites 
(Here desymmetrization is performed also for spin reversal); 
$1282 \times 1282$ for $S^z_{\rm tot}=1$  and $K_{\rm tot}=0$. 

Numerical calculations are performed for the $XY$-like region,
$|\Delta|<1$, where $\Delta$ is the anisotropic parameter $\Delta_1$ or
$\Delta_2$. 
It may be interesting to study for the Ising-like region,
$|\Delta|>1$, because there exist Ising-like magnets.
For example, CsCoBr$_3$ and CsCoCl$_3$ are quasi-1D Ising-like 
antiferromagnets with $\Delta\sim 10$.~\cite{Nagler} For $\Delta\gg 1$,
however, level statistics is not reliable because energy spectra have
some large gaps relative to $\Delta$ and the above unfolding method 
is invalid.

%
%
\section{\label{sec:NNN} Next-Nearest-Neighbor Coupled XXZ Spin Chains}

We now discuss numerically in the section that for the XXZ spin chains  
the characteristic behavior 
of level statistics is determined through competition among 
quantum chaos, quantum integrability and finite-size effects.

\subsection{\label{sec:2SR} Spin reversal symmetry and Wigner behavior 
for $S_{tot}^{z}=0$ }

\begin{figure}
\begin{center}
\includegraphics[width=8cm]{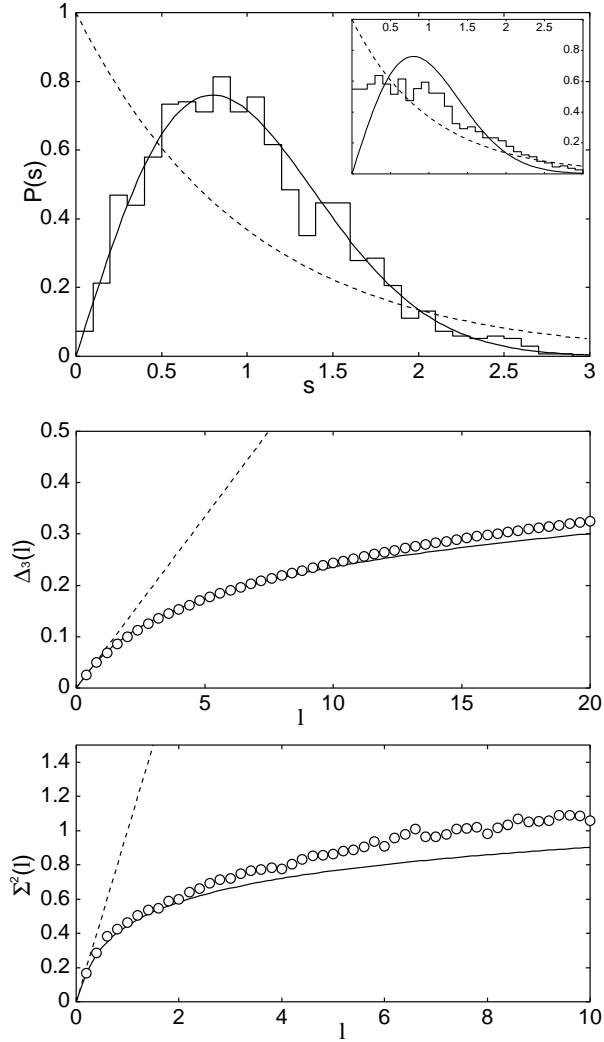}
\end{center}
\caption{\label{fig:nph} Level-spacing distribution $P(s)$, spectral
 rigidity $\Delta_3(l)$, and number variance $\Sigma^2(l)$ of the NNN
 coupled chain for $L=18$, $J_2/J_1=0.5$, $\Delta_1=\Delta_2=0.5$,   
 $S^z_{\rm tot}=0$, $K_{\rm tot}=2\pi/L$ under complete
 desymmetrization.  
Broken lines correspond to Poisson behavior, and 
 solid curves Wigner behavior.
The inset shows $P(s)$ of the same system under incomplete
 desymmetrization. Namely,  
the desymmetrization with respect to spin
reversal symmetry is not performed for the inset.}
\end{figure}

 For a sector of $S^z_{\rm tot}=0$ 
we numerically discuss the characteristic behavior of level statistics 
on the XXZ spin chains. 
Here we note that spin reversal symmetry 
has not been considered explicitly 
in previous studies of level statistics for various quantum spin chains. 
In some sense, desymmetrizing the Hamiltonian with respect to   
spin reversal symmetry has been avoided due to some technical 
difficulty. Level statistics has been discussed 
only for sectors of $S^z_{\rm tot} \ne 0$, where 
there is no need of the desymmetrization with respect to 
spin reversal symmetry.

Let us show explicitly such a case that Wigner behavior appears 
for $S^z_{\rm tot}=0$  
if we consider spin reversal symmetry. 
In Fig.~\ref{fig:nph} we have obtained the numerical results 
for level statistics such as the level-spacing distribution $P(s)$,  
the spectral rigidity $\Delta_3(l)$ and the number variance $\Sigma^2(l)$,  
for the sector of $K_{\rm tot}=2 \pi/L$ and $S_{\rm tot}^{z}=0$. 
Here we note that in the sector  
parity invariance does not exist and 
 we focus on spin reversal symmetry. 
The numerical results of  level statistics shown in  Fig.~\ref{fig:nph} 
clearly suggest Wigner behavior.
The curve of the Wigner distribution 
fits well to the data of the level-spacing distribution $P(s)$ in the
main panel. 
The plots of the spectral rigidity $\Delta_3(l)$ 
 are consistent with the curve of Wigner behavior especially for small $l$  
as shown in Fig.~\ref{fig:nph}. 
It is also the case with the number variance $\Sigma^2(l)$.
We see small deviations of  $\Delta_3(l)$ and $\Sigma^2(l)$ for large $l$
because of some finite-size effects. 
We have thus confirmed that Wigner behavior 
appears also in the sector of $S_{\rm tot}^z=0$ for 
the XXZ spin chains with the NNN interaction. 

In the inset of Fig.~\ref{fig:nph}, we have shown the 
level-spacing distribution for which we do not
perform the desymmetrization with respect to spin reversal.   
It shows non-Wigner behavior, which is similar to that of some other 
cases of mixed symmetry as we shall show in Fig.~\ref{fig:Sz}.

 The spin reversal operation on the spin variable of the 
$j$th site is defined by 
\begin{equation}
 S^{\pm}_j \to S^{\mp}_j, \quad S^z_j \to -S^z_j.
\label{eq:reversal}
\end{equation} 
Here, $S^{\pm}_{j} =(S^x_j \pm {\rm i} S^y_j)/2$. 
Let $M$ denote the number of down-spins in a given sector.  
The value of the total spin operator $S_{\rm tot}^z$ is given by 
$S_{\rm tot}^z=L/2-M$. 
Some details of spin reversal operation are given in 
Appendices~\ref{sec:SR} and \ref{sec:vc}.

\subsection{\label{sec:2FSE} Finite-size effects on the level-spacing
distribution} 

Let us explicitly discuss finite-size effects appearing 
in level statistics. 
They are important 
in the Poisson-like or non-Wigner
behavior observed in level statistics 
for the completely desymmetrized XXZ Hamiltonians. 
There are two regions in which finite-size effects 
are prominent: A region 
where $J_2$ is close to zero and  
 another region where $\Delta_1$ and $\Delta_2$ are close to 1. 
 In the former region  quantum integrability appears 
 through finite-size effects, and  
the characteristic behavior of level statistics becomes close to  
Poisson-like behavior.    
In the latter region,  Poisson-like behavior appears  
 due to the symmetry enhancement at the point 
of $\Delta_1=\Delta_2=1$, where the $U(1)$ symmetry 
of the XXZ spin chain expands into the spin $SU(2)$ symmetry.

\begin{figure*}
\begin{center}
\includegraphics[width=14cm]{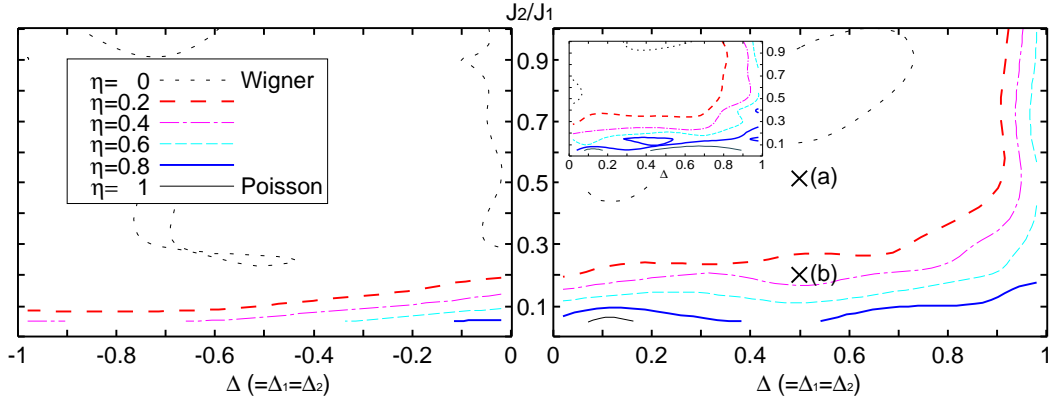}
\end{center}
\caption{\label{fig:diag} The diagram of contour lines of $\eta$ 
for the NNN coupled  chain with $L=18$ (in the inset, $L=16$), 
in the sector of $S^z_{\rm tot}=1$ and $K_{\rm tot}=0$. 
 Roughly speaking, 
 the area above the thick long-dashed line, Wigner behavior; 
 the area below the thick solid line, Poisson behavior. 
 The points (a) and (b) correspond to
 Figs.~\ref{fig:nregion}(a) and \ref{fig:nregion}(b), respectively.}
\end{figure*}

Let us now discuss how the degree of non-Wigner behavior depends on  
the anisotropy parameters, $\Delta_1$ and $\Delta_2$, 
 and the NNN coupling, $J_2$. For simplicity we set 
 $\Delta_1=\Delta_2$ and denote it by $\Delta$, and we also consider   
the ratio of  $J_2/J_1$. 
We express the degree of non-Wigner behavior by the following parameter: 
\begin{equation}
 \eta=\frac{\int^{s_0}_0[P(s)-P_{\rm Wig}(s)]ds}
{\int^{s_0}_0[P_{\rm Poi}(s)-P_{\rm Wig}(s)]ds},
\label{eq:eta}
\end{equation} 
where $s_0=0.4729\cdots$ is the intersection point of $P_{\rm Poi}(s)$
and $P_{\rm Wig}(s)$.~\cite{Georgeot,random} We have $\eta=0$ when $P(s)$
coincides with $P_{\rm Wig}(s)$, and $\eta=1$ when $P(s)$ coincides with  
$P_{\rm Poi}(s)$. The diagram of contour lines of $\eta$ is shown in 
Fig.~\ref{fig:diag}. We have calculated them for  
the area $-0.98 \le \Delta \le
-0.02$, $0.02 \le \Delta \le 0.98$, and $0.02 \le J_2/J_1 \le 1$, where 
$\Delta=\Delta_1=\Delta_2$.

The contour lines of $\eta$ show that 
a behavior close to Wigner one 
appears in a large region, while the 
Poisson-like behavior appears in a narrow region 
along the line of $J_2/J_1=0$ and that of $\Delta_1 =\Delta_2=1$. 
The Poisson-like behavior is dominated by finite-size effects 
and hence should  vanish when $L\to\infty$. This expectation is  
consistent with the suggestion in ref.~\citen{rabson} that an
infinitesimal integrability-breaking term (the NNN term of 
eq.~(\ref{eq:NNN}) in this paper) would lead to Wigner behavior. 
In fact, it is seen in the inset of Fig.~\ref{fig:diag} 
that the region of Wigner behavior shrinks for $L=16$. 
Here we remark that the phase diagrams of the ground
state~\cite{HN,Nomura} are totally different from the diagram of contour
lines of $\eta$. It is due to the fact that level statistics reflects 
highly excited states rather than the ground state.

When $\Delta_1=\Delta_2=0.98$,  
Poisson-like behavior appears in level statistics 
due to some finite-size effects. It will be explicitly   
shown in Figs.~\ref{fig:Sz} and \ref{fig:rid1}.
When  $\Delta_1=\Delta_2=1$, 
eq.~(\ref{eq:NNN}) coincides with the Heisenberg chain, 
which has the spin $SU(2)$ symmetry. 
Some degenerate energy levels at $\Delta_1=\Delta_2=1$  
can become nondegenerate when $\Delta_1$ and $\Delta_2$ are not equal to 1.  
The difference among the nondegenerate energy levels 
should be smaller than the typical level spacing 
when $\Delta_1$ and $\Delta_2$ are close to 1.  
The typical level spacing, which is of the order of $1/L$, should become 
large when the system size $L$ is small. 
Thus, the Poisson-like behavior should practically 
appear in level statistics. 
We note that for the Heisenberg chain  
Wigner behavior appears 
in the level-spacing distribution 
when we desymmetrize the Hamiltonian 
with respect to the spin $SU(2)$ symmetry.~\cite{Hsu,Poil}

\subsection{Homogeneity of the characteristic behavior of 
level statistics throughout the spectrum} 

\begin{figure}
\begin{center}
\includegraphics[width=8cm]{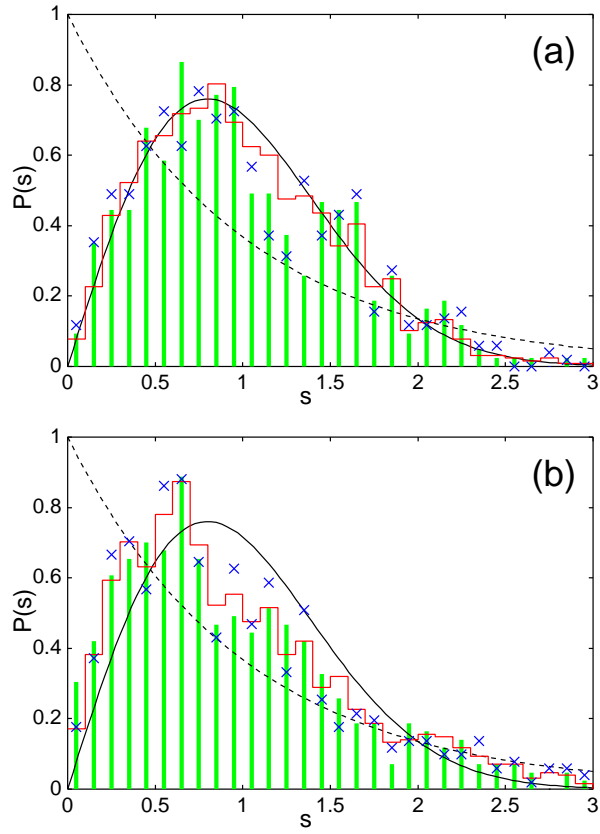}
\end{center}
\caption{\label{fig:nregion} Level-spacing distribution of the NNN
 coupled chain for $L=18$, $\Delta_1=\Delta_2=0.5$, $S^z_{\rm tot}=1$,
 and (a) $J_2/J_1=0.5$; (b) $J_2/J_1=0.2$. 
 Red histograms are for all levels; green bars, $1/3$ of
 all levels around the center; blue crosses, 10\% of all levels from 
 each of the two edges. Solid and broken lines show the Wigner and Poisson 
 distributions, respectively. }
\end{figure}

Let us discuss that for the XXZ spin chains 
the characteristic behavior of level statistics 
does not depend on the energy range of the spectrum. 
In Fig.~\ref{fig:nregion}, we show level-spacing distributions 
evaluated at points (a) and (b) shown in the diagram of 
Fig.~\ref{fig:diag}. They are evaluated for three different energy ranges. 
The distributions shown in 
Fig.~\ref{fig:nregion}(a) give Wigner behavior, while the distributions 
of Fig.~\ref{fig:nregion}(b) are close to Poisson behavior.

Let us explain the three different 
energy ranges shown in Fig.~\ref{fig:nregion}. 
Histograms show the level-spacing distributions evaluated 
for all levels, while bars show those evaluated only 
for the $1/3$ of all levels around the center,  
and crosses for the 10\% of all levels located from    
 each of the two spectral edges.

Quite interestingly, 
the distributions evaluated for the different energy ranges are 
quite similar to each other. 
It should be typical of frustrated quantum systems.
In frustrated quantum systems, quantum chaotic behavior appears already
in the low energy region near the ground state.~\cite{Nakamura}
For non-frustrated cases, however,
level statistics shows Poisson-like behavior
in a low region. In ref.~\citen{Zotos}, for example, level statistics is
Poisson-like for the low energy spectrum of one-dimensional
 spinless fermions with
the nearest hopping, the nearest- and next-nearest-neighbor interactions
in a $1/3$-filling case. The model is related to our model by a
transformation, while, the $1/3$-filling case is not a frustrated case.
Furthermore, there is another example of Poisson behavior. The level
statistics of the Anderson model shows Poisson behavior even in the
metallic phase, if we evaluate it in the edge regions of energy spectrum.

\subsection{Two solutions to unexpected non-Wigner behavior: 
mixed symmetry and finite-size effects}

Unexpected non-Wigner behavior has been reported 
in ref.~\citen{NNN} for level-spacing 
distributions of the NNN coupled XXZ chains. 
Let us discuss the reason why  it was observed, 
considering both mixed symmetry and finite-size effects. 
Here we explain mixed symmetry in the following: Suppose that the
Hamiltonian of a system is invariant under symmetry operations $T_j$ for
$j=1,2,\ldots , m$, which are commuting each other. The
eigenstates of the Hamiltonian are specified
by the set of the eigenvalues of all $T_j$. If we desymmetrize
the spectrum with respect to a partial set of symmetry operations,
we say that the derived spectrum has mixed symmetry. For instance, if we
only consider symmetry operations $T_j$ for $j=1,2,\ldots ,m-1$, then
the contributions of such eigenstates with different eigenvalues of
$T_m$ can be mixed in the derived spectrum.

There are two types of non-Wigner profiles reported in ref.~\citen{NNN}
for the nonintegrable systems: 
 one is  given by almost the numerical average of the 
Poisson and the Wigner distributions,  
  and another one is rather close to the Poisson distribution. 
The profiles of the first type appear 
in various cases,~\cite{NNN} such as the case of $\Delta_2=0.5$.  
The profiles of the second type appear 
in particular for 
the case of $\Delta_1\simeq \Delta_2 \simeq 1$. We may call the latter 
Poisson-like behavior rather than simple non-Wigner behavior. 
Both types of non-Wigner distributions 
have been observed in the subspace of $S^z_{\rm tot}=0$, 
which is the largest sector of the 
Hamiltonian matrix of eq.~(\ref{eq:NNN}). 
Here we note that the observations in ref.~\citen{NNN} are obtained
particularly for $S^z_{\rm tot} = 0$, while in ref.~\citen{rabson} the
Wigner behavior is observed for $S^z_{\rm tot} \ne 0$ in the
level-spacing distributions of similar 
XXZ chains.

\begin{figure*}
\begin{center}
\includegraphics[width=14cm]{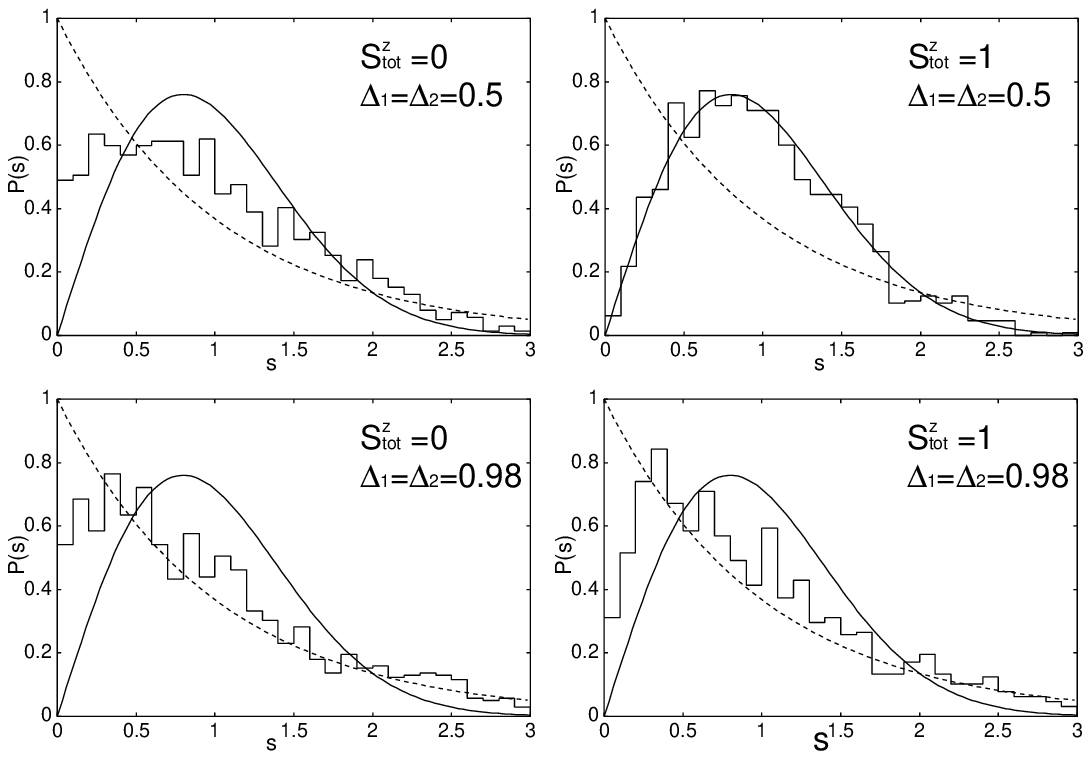}
\end{center}
\caption{\label{fig:Sz} Level-spacing distribution $P(s)$ of the NNN
 coupled chain for $L=18$, $J_2/J_1=1$, $K_{\rm tot}=0$. Broken lines, 
the Poisson
 distribution; solid curves, the Wigner distribution.}
\end{figure*}

We give level-spacing distributions in Fig.~\ref{fig:Sz} 
for the four cases:  $S^z_{\rm tot}=0$ or $1$ 
and $\Delta_1=\Delta_2=0.5$ or $0.98$. 
The numerical results suggest 
that the value of $S^z_{\rm tot}$ should be important 
as well as the anisotropy parameters, $\Delta_1$ and $\Delta_2$, 
in the observed non-Wigner behavior of the level-spacing distributions. 
When $\Delta_1=\Delta_2=0.5$, 
Wigner behavior appears for $S^z_{\rm tot}=1$,   
while the non-Wigner behavior was observed for $S^z_{\rm tot}=0$.   
We have also checked that Wigner behavior appears for $S^z_{\rm tot}=2$.
Furthermore, we have confirmed 
that such $S^z_{\rm tot}$-dependence of the level-spacing 
distribution is valid for some values of $K_{\rm tot}$.
Here we have desymmetrized the Hamiltonian 
according to $S^z_{\rm tot}$, $K_{\rm tot}$, and the parity 
when it exists, but not to the spin reversal.  
Here we note that the parity invariance exists only for sectors with 
$K_{\rm tot}=0$ or $\pi$ when $L$ is even.

The non-Wigner behavior observed for the case $S^z_{\rm tot}=0$ and 
$\Delta_1=\Delta_2=0.5$ shown in Fig.~\ref{fig:Sz} is 
due to mixed symmetry. 
Let us recall that the system is invariant under spin rotation around
the $z$-axis, translation and reflection (parity). For 
$S^z_{\rm tot}=0$, the system is also invariant under another
operation, spin reversal.
Here, the spectrum shown in Fig.~\ref{fig:Sz} has not been desymmetrized
with respect to the spin reversal symmetry due to some technical
aspect.~\cite{note}    
However, we expect that Wigner behavior
will appear for the case $S_{\rm tot}=0$ and $\Delta_1=\Delta_2=0.5$, if
we further perform
desymmetrization with respect to spin reversal symmetry. There are
two reasons supporting it:
First, for $K_{\rm tot}=2\pi /L$, Wigner behavior for $S^z_{\rm tot}=0$
has been shown in Fig.~\ref{fig:nph}.
Here the system is invariant under spin rotation around
the $z$-axis, translation and spin reversal, but not for reflection. 
Secondly, the behavior of level statistics is independent of $K_{\rm tot}$
for a large number of cases we have investigated. 
For example, the non-Winger behavior for the case $S^z_{\rm tot}=0$ and 
$\Delta_1=\Delta_2=0.5$ is very similar to that of the inset of 
Fig.~\ref{fig:nph}. 

The Poisson-like 
behavior for the case $\Delta_1=\Delta_2=0.98$ 
is dominated by finite-size effects. In fact, 
for $S^z_{\rm tot}=1$ of Fig.~\ref{fig:Sz}, 
the Poisson-like behavior appears when $\Delta_1=\Delta_2=0.98$, 
while Wigner behavior appears when $\Delta_1=\Delta_2=0.5$. 
We have confirmed that such tendency does not depend on the value of 
$K_{\rm tot}$:  we see it not only 
for $K_{\rm tot}= 0$ but also for $K_{\rm tot} \ne 0$.

\begin{figure}
\begin{center}
\includegraphics[width=8cm]{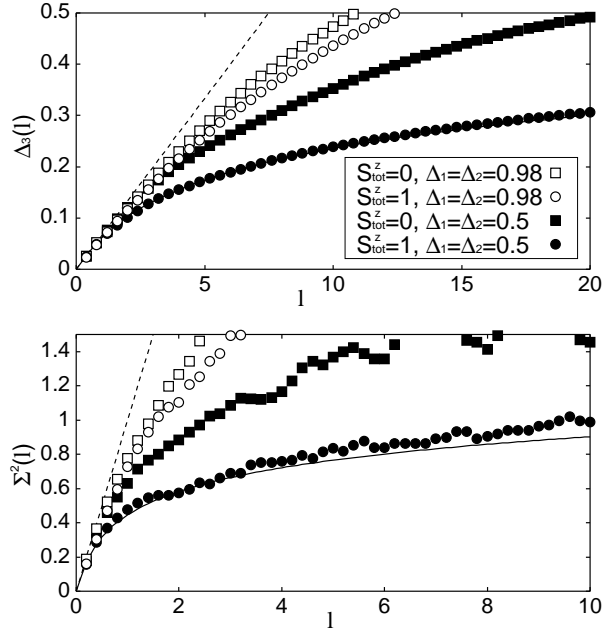}	       
\end{center}
\caption{\label{fig:rid1} Spectral rigidity $\Delta_3(l)$ and number
 variance $\Sigma^2(l)$ of the NNN coupled chain for $L=18$, $J_2/J_1=1$, 
$K_{\rm tot}=0$. Broken
 lines, Poisson behavior; Solid lines, Wigner behavior. In each of 
 the four distributions $P(s)$ vanishes at $s=0$: 
 there is no degeneracy among energy levels. }
\end{figure}

The observations of the level-spacing distributions 
can also be confirmed by investigating 
 spectral rigidity $\Delta_3(l)$ and 
number variance $\Sigma^2(l)$.  
In Fig.~\ref{fig:rid1}, $\Delta_3(l)$ and $\Sigma^2(l)$ are shown
for the four cases 
corresponding to those of Fig.~\ref{fig:Sz}. For 
$S^z_{\rm tot}=1$ and $\Delta_1=\Delta_2=0.5$, Wigner behavior appears. 
 For $S^z_{\rm tot}=0$ and $\Delta_1=\Delta_2=0.5$, 
 an intermediate behavior appears, which is close to 
 the average between Wigner and Poisson behaviors. 
For  $\Delta_1=\Delta_2=0.98$, 
both for $S^z_{\rm tot}=0$ and $S^z_{\rm tot}=1$, 
Poisson-like behavior appears for $\Delta_3(l)$ and $\Sigma^2(l)$.

\section{\label{sec:integrable} Integrable XXZ Spin Chain in a
special case}

\begin{figure}
\begin{center}
\includegraphics[width=8cm]{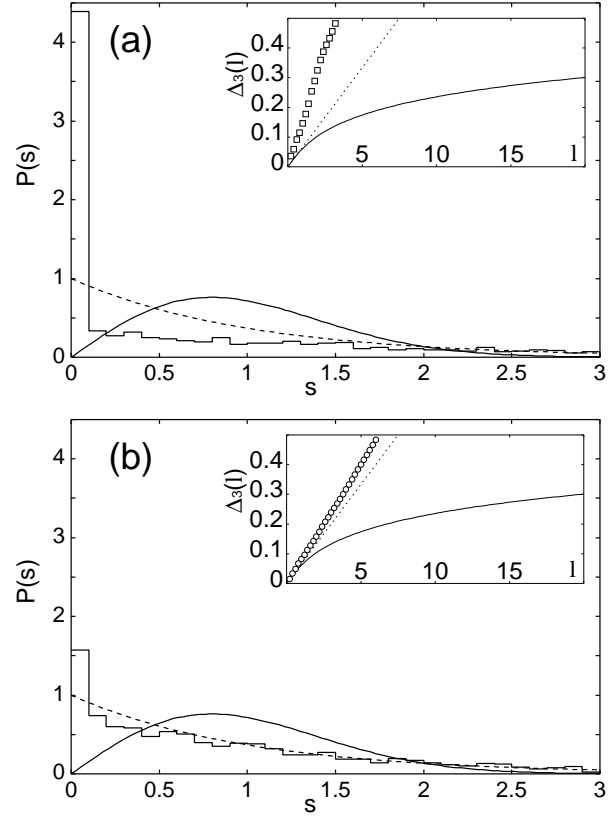}
\end{center}
\caption{\label{fig:super} Level-spacing distribution $P(s)$ of the 
 integrable chain ($J_2=0$) for $L=18$, $\Delta_1=0.5$, 
$S^z_{\rm tot}=0$, $K_{\rm tot}=2\pi/L$. 
(a) Desymmetrization with respect to spin reversal is not performed. 
(b) Desymmetrization with respect to spin reversal is performed.
The inset is  spectral rigidity for each case.}
\end{figure}

Let us discuss level statistics for the special  case 
of the integrable XXZ spin chain that has  
the $sl_2$ loop algebra symmetry. Here we recall that 
the XXZ spin chain is integrable when the NNN coupling $J_2$ vanishes, 
and also that the $sl_2$ loop algebra symmetry exists when 
$q$ is a root of unity.   
Here the anisotropy $\Delta_1$ is related to 
$q$ through $\Delta_1=(q+q^{-1})/2$.  
For instance, when the parameter $q$ is given by $\exp({\rm i} \pi/3)$, 
we have $\Delta=0.5$.

The level spacing distribution  $P(s)$ and the spectral rigidity 
$\Delta_3(l)$ of the integrable XXZ spin chain  
are shown for $\Delta_1=0.5$ in Fig.~\ref{fig:super}. 
In Fig.~\ref{fig:super} (a) we do not perform the desymmetrization
according to spin reversal, while  
in Fig.~\ref{fig:super} (b) we plot  
the level-spacing distribution  $P(s)$ and the spectral rigidity 
$\Delta_3(l)$ after we desymmetrize the Hamiltonian 
with respect to the spin reversal operation.

We observe that there still remain many degeneracies 
associated with the $sl_2$ loop algebra symmetry,  
even after desymmetrizing the Hamiltonian 
with respect to  spin reversal symmetry.  
The level-spacing distribution $P(s)$ has a small peak at 
$s=0$ in Fig.~\ref{fig:super} (b). 
Furthermore, the slopes of $\Delta_3(l)$ shown in the insets of 
Figs.~\ref{fig:super} (a) and \ref{fig:super} (b) 
are larger than that of Poisson behavior. 
However, the numerical result does not necessarily give a counterexample 
to the conjecture of RMT. The level statistics might show Poisson behavior,   
if we completely desymmetrize the Hamiltonian matrix 
in terms of the $sl_2$ loop algebra symmetry.

\section{\label{sec:conclusions} Conclusions}  

For the finite spin-$\frac12$ XXZ spin chains 
with the NNN interaction, 
we have evaluated characteristic quantities of level statistics 
such as the level-spacing distribution, the spectral rigidity and 
the number variance. 
 Through the numerical results we have obtained the following conjecture: 
When the symmetry of a finite-size system enhances at some point 
of the parameter space, 
the characteristic behavior of level statistics should 
be given by Poisson-like behavior near some region close to the point.  
In particular, 
we have shown that finite-size effects play an important  
role in the characteristic quantities of level statistics 
for the XXZ spin chains. Here they are integrable for $J_2=0$, and 
their $U(1)$ symmetry extends into $SU(2)$ symmetry 
at the point of $\Delta_1=\Delta_2=1$. 
Furthermore,   we have also shown that 
some unexpected non-Wigner behavior appears when an extra symmetry
exists and is not considered for desymmetrization, 
such as the case of the spin reversal symmetry in the 
sector of $S^{z}_{\rm tot}=0$.  
Here we note that in some cases
 extra symmetries depend on some parameters or
  some quantum numbers. We have thus solved completely 
the observed non-Wigner behaviors of NNN coupled chain 
for $S^z_{\rm tot}=0$ in ref.~\citen{NNN}. 

\section*{Acknowledgement} 

The authors would like to thank K.~Nakamura and T.~Kato for useful
 discussions. The authors also thank the Yukawa Institute for
 Theoretical Physics at Kyoto University. Discussions during the YITP
 workshop YITP-W-03-16 on ``Quantum Mechanics and Chaos: From
 Fundamental Problems through Nanosciences'' were useful to complete
 this work.
 The present study was partially supported by the Grant-in-Aid for
 Encouragement of Young Scientists (A): No.~14702012. It was also
 partially supported by Hayashi Memorial Foundation for Female Natural
 Scientists. 

\appendix

\section{\label{sec:JW} Jordan-Wigner and Fourier Transformations}

Let us rewrite eq.~(\ref{eq:NNN}) by
\begin{eqnarray}
&\mathcal{H}&=\mathcal{H}_1 + \mathcal{H}_2 \nonumber \\
&&=J_1 \sum^L_{l=1}(S^x_l S^x_{l+1} + S^y_l S^y_{l+1}
 + \Delta_1 S^z_l S^z_{l+1}) 
 + J_2 \sum^L_{l=1}(S^x_l S^x_{l+2} + S^y_l S^y_{l+2}
 + \Delta_2 S^z_l S^z_{l+2}),
\end{eqnarray}
where $\mathcal{H}_1$ is the term containing nearest-neighbor couplings
and $\mathcal{H}_2$ is the term containing next-nearest-neighbor
couplings. We define the Jordan-Wigner transformation by
 \begin{eqnarray}
 \sigma^{-}_l&=&\exp\left( -{\rm i}\pi \sum _{j=1}^{l-1} c_j^{\dagger} c_j
  \right)
 c_l^{\dagger},\label{eq:JWta}\\
 \sigma^{+}_l&=&\exp\left( {\rm i}\pi \sum _{j=1}^{l-1} c_j^{\dagger} c_j 
 \right)
 c_l, \label{eq:JWtb}\\
 \sigma^z_l&=& 2\sigma^{+}_l \sigma^{-}_l -1 = 
 1- 2 \sigma^{-}_l \sigma^{+}_l,\label{eq:JWtc}
 \end{eqnarray}
where $c^{\dagger}_j$ and $c_j$ are the creation and annihilation
operators of fermions on $j$th site. Under the Jordan-Wigner
transformation, $\mathcal{H}_1$ is written by
\begin{eqnarray}
 \mathcal{H}_1 &=& \frac{J_1}4 \Delta_1 
\left( L-2\sum_{l=1}^L c_l^{\dagger} c_l\right) \nonumber \\
&+& \frac{J_1}4 \sum_{l=1}^{L-1}
[ 2(c_l^{\dagger} c_{l+1} + c_{l+1}^{\dagger}c_l ) 
+\Delta_1 (4c_l^{\dagger}c_l c_{l+1}^{\dagger}c_{l+1} 
-2c_{l+1}^{\dagger}c_{l+1})]  \nonumber \\
&+& \frac{J_1}4 [ 2(c_L^{\dagger}c_{L+1} + c_{L+1}^{\dagger}c_L)
+\Delta_1(4c_L^{\dagger}c_L c_{L+1}^{\dagger}c_{L+1} 
-2c_{L+1}^{\dagger}c_{L+1})].
\end{eqnarray} 
Here, we consider periodic boundary conditions ($\sigma_{L+1}^{\pm} = 
\sigma_1^{\pm}$):
\begin{eqnarray}
 c_{L+1} &=& \exp\left( -{\rm i}\pi \sum_{j=1}^L \sigma_j^{-}\sigma_j^{+} 
 \right) \sigma_{L+1}^{+} 
 = \exp\left( -\frac{{\rm i}\pi L}2 + \frac{{\rm i}\pi}2 \sum_{j=1}^L 
 \sigma^z_j 
 \right) \sigma_1^{+}\nonumber \\
 &=&  \sigma_1^{+} \exp\left[ {\rm i}\pi\left( -\frac{L}2 + S^z_{\rm tot} +1
 \right) \right],\nonumber \\
 c_{L+1}^{\dagger} &=& 
\exp\left( {\rm i}\pi \sum_{j=1}^L \sigma_j^{-}\sigma_j^{+} 
 \right) \sigma_{L+1}^{-}
 = \exp\left( \frac{{\rm i}\pi L}2 - \frac{{\rm i}\pi}2 \sum_{j=1}^L 
 \sigma^z_j 
 \right) \sigma_1^{-}\nonumber \\
 &=&  \sigma_1^{-} \exp\left[ {\rm i}\pi\left( \frac{L}2 - S^z_{\rm tot} +1
 \right) \right],
\end{eqnarray}
where we use ${\rm e}^{S_1^z} \sigma_1^{\pm}=\sigma_1^{\pm} {\rm e}^{S_1^z\pm
1}$. Since $\sigma_1^{+}=c_1$, $\sigma_1^{-}=c_1^{\dagger}$ , and
$S^z_{\rm tot}=(L-M)/2-M/2=L/2-M$, where $M$ is the number of fermions, 
we have 
\begin{equation}
 c_{L+1}=-(-1)^M c_1, \quad c_{L+1}^{\dagger}=-(-1)^M c_1^{\dagger}.
\end{equation}
Therefore, $\mathcal{H}_1$ is rewritten as
\begin{eqnarray}
 \mathcal{H}_1 &=&\frac{J_1}4 \Delta_1 \left[ L +\sum_{l=1}^L
+(4c_l^{\dagger} c_l c_{l+1}^{\dagger}c_{l+1} -4c_l^{\dagger} c_l) 
\right]\nonumber \\
&+&\frac{J_1}4 \cdot 2 \left[ \sum_{l=1}^{L-1}
(c_l^{\dagger}c_{l+1} + c_{l+1}^{\dagger}c_l) -(-1)^M
(c_L^{\dagger}c_1 +c_1^{\dagger}c_L) \right].
\end{eqnarray}

Now, we define the Fourier transformation as
\begin{eqnarray}
 c_l={\frac 1 {\sqrt{L}}} \sum_k {\hat c}_{k} {\rm e}^{{\rm i} kl},
\label{eq:Fta} \\
c_l^{\dagger}={\frac 1 {\sqrt{L}}} \sum_k {\hat c}_{k}^{\dagger} 
{\rm e}^{-{\rm i} kl}.
\label{eq:Ftb}
\end{eqnarray} 
Here, $k$ takes $(2\pi /L)\times (\mbox{\rm an integer})$ 
for odd $M$ and $(2\pi /L)\times (\mbox{\rm a half-integer})$ for even $M$,
and $0 \le k < 2\pi$. After the Fourier transformation we have,
\begin{eqnarray}
 \mathcal{H}_1 &=& \frac{J_1}4 \Delta_1 L +J_1\sum_k
( \cos k -\Delta_1) {\hat c}_{k}^{\dagger} {\hat c}_{k} \nonumber \\
&-& \frac{J_1 \Delta_1}L \sum_{k_1,k_2,k_3,k_4} \delta_{k_1+k_2,k_3+k_4}
\exp\left[- {\rm i} (k_2-k_4) \right] 
{\hat c}_{k_1}^{\dagger}{\hat c}_{k_2}^{\dagger} {\hat c}_{k_3} {\hat c}_{k_4},
\label{eq:A1}
\end{eqnarray}
where
\begin{equation}
 \delta_{k_1+k_2,k_3+k_4}=\left\{
  \begin{array}{ll}
 1 & \quad \mbox{when $k_1+k_2$(mod $L$)=$k_3+k_4$(mod $L$)}.\\
 0 & \quad \mbox{otherwise}.
  \end{array}\right.
\end{equation}
Considering combination of $k$'s, we rewrite eq.~(\ref{eq:A1}) as
\begin{eqnarray}
\mathcal{H}_1&=&J_1 \Delta_1\left(\frac{L}4 -M \right) 
+J_1\sum_k \cos k {\hat c}_{k}^{\dagger}{\hat c}_{k} \nonumber\\
&-&\frac{2J_1\Delta_1}L \sum_{k_1 <k_2 \atop k_3>k_4}
\delta_{k_1+k_2,k_3+k_4}
\left[\cos \left(k_2-k_4\right)
-\cos\left(k_2-k_3\right)\right]
{\hat c}_{k_1}^{\dagger} {\hat c}_{k_2}^{\dagger} 
{\hat c}_{k_3} {\hat c}_{k_4}.
\end{eqnarray}

In the same way, $\mathcal{H}_2$ is rewritten as
\begin{eqnarray}
\mathcal{H}_2 &=& J_2 \Delta_2\left(\frac{L}4 -M \right) 
+J_2\sum_k \cos 2k {\hat c}_{k}^{\dagger}{\hat c}_{k} \nonumber\\
 &+& \frac{2J_2}L \sum_{k_1 <k_2 \atop k_3>k_4} \delta_{k_1+k_2,k_3+k_4}
\left[
\cos(k_1+k_3)
+\cos(k_2+k_4) 
- \cos(k_1+k_4)-\cos(k_2+k_3)\right]
{\hat c}_{k_1}^{\dagger}{\hat c}_{k_2}^{\dagger}
{\hat c}_{k_3}{\hat c}_{k_4} \nonumber\\
 &-& \frac{2J_2\Delta_2}L 
\sum_{k_1 <k_2 \atop k_3>k_4} \delta_{k_1+k_2,k_3+k_4}
\left\{\cos\left[2\left(k_2-k_4\right)\right]
-\cos\left[2\left(k_2-k_3\right)\right]\right\} 
{\hat c}_{k_1}^{\dagger}{\hat c}_{k_2}^{\dagger}
{\hat c}_{k_3}{\hat c}_{k_4}. 
\end{eqnarray}

\section{\label{sec:SR} Spin reversal symmetry on momentum-based fermions}

Let us find the momentum-based expression of the mapping 
 corresponding to the spin reversal transformation 
($S_j^{\pm} \to S_j^{\mp}$, $S_j^z \to -S_j^z$). According to
eqs.~(\ref{eq:JWta}), (\ref{eq:JWtb}), (\ref{eq:Fta}), and (\ref{eq:Ftb}),
\begin{eqnarray}
 {\hat c}_{k} = \frac1{\sqrt{L}} \sum_{l=1}^L \exp(-{\rm i} kl)\exp\left( 
-{\rm i}\pi\sum_{j=1}^{l-1}\sigma_j^{-}\sigma_j^{+} \right) \sigma_l^{+},
\label{eq:C1} \\ 
 {\hat c}_{k}^{\dagger}=\frac1{\sqrt{L}}\sum_{l=1}^L \exp({\rm i} kl)
\exp\left( 
{\rm i}\pi\sum_{j=1}^{l-1}\sigma_j^{-}\sigma_j^{+} \right) \sigma_l^{-}.
\label{eq:C1+} 
\end{eqnarray}
Under the transformation $\sigma_l^{\pm}\to\sigma_l^{\mp}$,
eqs.~(\ref{eq:C1}) and (\ref{eq:C1+}) is transformed as follows.
\begin{eqnarray}
 {\hat c}_{k} &\to & \frac1{\sqrt{L}}\sum_{l=1}^L \exp(-{\rm i} kl)\exp\left( 
-{\rm i}\pi\sum_{j=1}^{l-1}\sigma_j^{+}\sigma_j^{-} \right) \sigma_l^{-}
\nonumber \\
&=& \frac1{\sqrt{L}}\sum_{l=1}^L\exp(-{\rm i} kl)\exp\left[ 
-{\rm i}\pi\sum_{j=1}^{l-1}(I_j-\sigma_j^{-}\sigma_j^{+}) \right] \sigma_l^{-}
\nonumber \\
&=& \frac1{\sqrt{L}}\sum_{l=1}^L\exp[-{\rm i} kl-{\rm i}\pi (l-1)]\exp\left( 
{\rm i}\pi\sum_{j=1}^{l-1}\sigma_j^{-}\sigma_j^{+} \right) \sigma_l^{-},
\label{eq:C2}
\end{eqnarray}
where $I_j$ is the unit matrix. Now, considering ${\rm e}^{2\pi{\rm i} l}=1$,
where $l$ is an integer, we find that
\begin{eqnarray}
 \exp[-{\rm i} kl-{ \rm i}\pi (l-1)]&=&
 \exp[-{\rm i} kl-{ \rm i}\pi (l-1)+2\pi{\rm i} l]
=\exp[{\rm i} (\pi-k)l+{ \rm i}\pi] \nonumber\\
&=& -\exp[{\rm i} (\pi-k)l].
\end{eqnarray}
Therefore, eq.~(\ref{eq:C2}) is rewritten as 
\begin{equation}
{\hat c}_{k} \to -\frac1{\sqrt{L}}\sum_{l=1}^L\exp[{\rm i} (\pi-k)l]
\exp\left( 
{\rm i}\pi\sum_{j=1}^{l-1}\sigma_j^{-}\sigma_j^{+} \right) \sigma_l^{-}
=-{\hat c}_{\pi-k}^{\dagger}.
\label{eq:ph}
\end{equation}
In the same way, we have 
\begin{equation}
{\hat c}_{k}^{\dagger} \to - {\hat c}_{\pi-k}.
\label{eq:ph1} 
\end{equation}

It is sometimes convenient to use the following 
in stead of eqs.~(\ref{eq:ph}) and (\ref{eq:ph1}): 
\begin{equation}
{\hat c}_k^{\dagger} \to {\hat c}_{\pi-k}, 
\quad {\hat c}_{k} \to c^{\dagger}_{\pi-k} . 
\label{eq:ph2}
\end{equation}
The Hamiltonians are invariant not only for eqs.~(\ref{eq:ph}) 
and (\ref{eq:ph1}) 
but also for eq~(\ref{eq:ph2}). The form (\ref{eq:ph2}) 
has an advantage that  
we do not need to consider the phase factor $-1$ that appears in 
eqs.~(\ref{eq:ph}) and (\ref{eq:ph1}).  

Making use of the spin reversal operation expressed 
in terms of  the fermion basis (\ref{eq:ph2}) 
we have desymmetrized the Hamiltonian matrix 
in the sector $S_{\rm tot}^{z}=0$ with respect to spin reversal 
symmetry. For a given vector with $S_{\rm tot}^{z}=0$  
we calculate how it transforms under the operation (\ref{eq:ph2}). If it 
is not a singlet and transforms into a different vector, 
then we combine the pair into an eigenvector of the operation (\ref{eq:ph2}).  

Similarly, when we desymmetrize the Hamiltonian with respect to parity,
we use the parity operation expressed in terms of the fermion basis:
\begin{equation}
{\hat c}_k^{\dagger} \to {\rm e}^{{\rm i}k} {\hat c}_{-k}^{\dagger} 
(-1)^{M-1 + L/2 - S^{z}_{\rm tot}} \, , \quad 
{\hat c}_k \to {\rm e}^{-{\rm i}k} {\hat c}_{-k} 
(-1)^{M +L/2 - S^{z}_{\rm tot}}. 
\label{eq:parity}
\end{equation}
Here we recall that $M$ denotes the number of down-spins in the sector 
of  $S_{\rm tot}^{z} = L/2 -M$. 
Through the Jordan-Wigner transformation we easily derive 
the parity operation (\ref{eq:parity}) from that on the spin variables: 
$\sigma_{l}^{\pm} \to \sigma_{L+1-l}^{\pm} \, , \quad 
\sigma_{l}^{z} \to \sigma_{L+1-l}^{z} \, $ for 
$l=1, 2, \ldots, L$.

\section{\label{sec:vc} Spin reversal operation on the vacuum state}

When desymmetrizing 
the Hamiltonian with respect to spin reversal symmetry, 
it is useful to know how the vacuum state transforms 
under the spin reversal operation expressed in terms of 
the momentum-based fermion operators. 
Let us denote by $| 0 \rangle$ the vacuum state where there is no down-spin.
Under the spin reversal operation 
it transforms up to a phase factor $A_L$ as follows 
\begin{equation}
| 0 \rangle \to A_L \,  
{\hat c}_{q_1}^{\dagger} {\hat c}_{q_2}^{\dagger} 
\cdots {\hat c}_{q_L}^{\dagger}  
| 0 \rangle 
\end{equation}
Here $q_j$  
 denotes  momentum $(2\pi/L)j$ for $j=1, 2, \ldots, L$, when $M$ is 
odd, and $(2\pi/L)(j-1/2)$ for $j=1, 2, \ldots, L$, when $M$ is even.  
The phase factor $A_L$ is given by 
\begin{equation}
  A_L = {\frac 1 {L^{L/2}}} \sum_{P \in {\cal S}_L } 
  \epsilon_P \exp(- {\rm i} \sum_{j=1}^{L} j k_{P j} ) \, .  
\label{eq:vacuum} 
  \end{equation}
Here ${\cal S}_L$ denotes the set of 
 permutations on $L$ elements, $\epsilon_P$ the sign of permutation $P$.   
 Furthermore, we can calculate the phase factor $A_L$  as follows 
 \begin{equation}
 A_L = \left\{  
 \begin{array}{ccc} 
 (-1)^\ell & {\rm for} & M=2 \ell \\
 (-1)^{\ell+1} & {\rm for} & M=2 \ell +1 \\
\end{array} 
            \right.
\label{eq:vacuum2}
 \end{equation}
The derivation is given in the following. 
Let us introduce the following matrix: 
\begin{equation}
U_{s} = \prod_{j=1}^{L} \sigma_j^x \, . 
\end{equation}
We may express the spin reversal operation (\ref{eq:reversal}) 
as follows  
\begin{eqnarray}
U_s  S^{\pm}_j U_s^{-1} & = & S^{\mp}_j \, , \quad  
U_s S^z_j U_s^{-1} = -S^z_j
\nonumber \\
U_s {\hat c}_{k}^{\dagger} U_s^{-1} & = & -{\hat c}_{\pi-k} \, , \quad 
U_s {\hat c}_{k} U_s^{-1} = -{\hat c}_{\pi-k}^{\dagger} .
\end{eqnarray}
We thus have 
\begin{equation} 
U_s \, | 0 \rangle = \prod_{j=1}^{L} \sigma_j^x \, | 0 \rangle = 
\sigma_1^{-}  \sigma_2^{-}  \cdots \sigma_L^{-} \, | 0 \rangle 
\end{equation} 
Applying the Jordan-Wigner transformation and   
substituting $c_{\ell}^{\dagger}$s with ${\hat c}_{k}^{\dagger}$s 
through (\ref{eq:Ftb}), we have 
\begin{eqnarray} 
\sigma_1^{-}  \sigma_2^{-}  \cdots \sigma_L^{-} \, | 0 \rangle 
 & = & c_1^{\dagger} c_2^{\dagger} \cdots c_L^{\dagger} | 0 \rangle  
\nonumber \\
 & = & {\frac 1 {L^{L/2}}} \sum_{k_1} \cdots \sum_{k_L} 
{\rm e}^{-{\rm i} (1 k_1 + 2 k_2 + \cdots + L k_L)} \, 
{\hat c}_{k_1}^{\dagger} {\hat c}_{k_2}^{\dagger} \cdots  
{\hat c}_{k_L}^{\dagger} | 0 \rangle \nonumber \\
& = & {\frac 1 {L^{L/2}}} \left( \sum_{P \in {\cal S}_L} 
{\rm e}^{(- {\rm i} \sum_{j=1}^{L} j k_{Pj} )} \, \epsilon_P \right) 
{\hat c}_{q_1}^{\dagger} {\hat c}_{q_2}^{\dagger} \cdots  
{\hat c}_{q_L}^{\dagger} | 0 \rangle 
\end{eqnarray}
Thus we have the expression (\ref{eq:vacuum}) of the phase factor $A_L$. 
Here we recall that $q_j= (2 \pi/L)j$ for $j=1, 2, \ldots, L$ for 
odd $M$, and  $q_j= (2 \pi/L)(j-1/2)$ for $j=1, 2, \ldots, L$ for 
even $M$. We also recall that $M$ denotes the number of down-spins 
in the sector.

We now calculate the expression (\ref{eq:vacuum2}) 
for the phase factor $A_L$. 
We take a vector $| v \rangle$ with $S_{tot}^{z}=0$ as follows.  
When $M$ is odd, we introduce $\ell=(M-1)/2$  
and we define $| v \rangle $ by    
$$ 
|v \rangle = \left({\hat c}_{1}^{\dagger} {\hat c}_{2}^{\dagger} 
 \cdots {\hat c}_{\ell}^{\dagger} \right) \cdot \left( 
 {\hat c}_{-\ell}^{\dagger}  
\cdots {\hat c}_{-2}^{\dagger} {\hat c}_{-1}^{\dagger} \right) \cdot 
{\hat c}_{0}^{\dagger}  | 0 \rangle
$$
Here ${\hat c}_{j}^{\dagger}$ denotes ${\hat c}_{k}^{\dagger}$ with 
$k= (2\pi/L) j$.  
When $M$ is even, we take $\ell=M/2$,  
and we define $| v \rangle $ by    
$$ 
|v \rangle = \left({\hat c}_{1/2}^{\dagger} {\hat c}_{3/2}^{\dagger} 
 \cdots  {\hat c}_{\ell-1/2}^{\dagger} \right) \cdot 
 \left( {\hat c}_{-(\ell-1/2)}^{\dagger}
  \cdots {\hat c}_{-3/2}^{\dagger} {\hat c}_{-1/2}^{\dagger} \right) 
  | 0 \rangle
$$
Here ${\hat c}_{j+1/2}^{\dagger}$ denotes ${\hat c}_{k}^{\dagger}$ with 
$k= (2\pi/L)(j+1/2)$.  
Through the operation (\ref{eq:ph}) we show that 
$U_s | v \rangle = (-1)^{\ell+1} A_L | v \rangle$ for 
$M$ odd and $U_s | v \rangle = (-1)^{\ell} A_L | v \rangle$ for 
$M$ even.  Thus, we have at least $U_s | v \rangle = \pm | v \rangle$.  

Let us show that $U_s | v \rangle = + | v \rangle$  for both odd $M$ and 
even $M$ cases. First we note that 
$U_s c_1^{\dagger} c_2^{\dagger} \cdots c_{L/2}^{\dagger} | 0 \rangle 
=   c_{L/2+1}^{\dagger} c_{L/2+2}^{\dagger} \cdots c_{L}^{\dagger} 
| 0 \rangle $.  Second, expanding the vector $| v \rangle$ in terms of 
$c_{j_1}^{\dagger} c_{j_2}^{\dagger} \cdots c_{j_{L/2}}^{\dagger} | 0 \rangle $ with $1 \le j_1 < j_2 < \cdots < j_{L/2} \le L$, we show that 
the coefficient of $c_1^{\dagger} c_2^{\dagger} \cdots c_{L/2}^{\dagger} 
| 0 \rangle$ in the expansion is equal to that of   
$c_{L/2+1}^{\dagger} c_{L/2+2}^{\dagger} \cdots c_{L}^{\dagger} | 0 \rangle $. 
Therefore we have $U_s | v \rangle = + |v \rangle$.  

Thus, we obtain the expression (\ref{eq:vacuum2}) 
for the phase factor $A_L$.


\begin{thebibliography}{99}
\bibitem{note} It is technically difficult to desymmetrize the
	 Hamiltonian with respect to both parity and spin reversal
	 symmetries simultaneously.
\end{thebibliography}
\end{document}